\documentclass[final]{elsart}   

\usepackage{graphicx}          

\usepackage{amssymb}

\providecommand{\lay}{{\rm lay}}
\providecommand{\pseudol}{\prec\!}
\providecommand{\pseudor}{\!\succ}

\begin{document}

\begin{frontmatter}



\title{{\bf Topological correlations and asymptotic freedom in cellular aggregates}}


\author[faez]{F. Miri},
\author[ch]{C. Oguey\corauthref{cor}}

\address[faez]{IASBS, Zanjan, Iran}
\ead{miri@iasbs.ac.ir}

\address[ch]{LPTM\thanksref{cnrs}, Univ. de Cergy-Pontoise, France}
\ead{oguey@ptm.u-cergy.fr}
\corauth[cor]{Corresponding author. Address: LPTM, Univ. de Cergy-Pontoise, 2, ave A. Chauvin, F--95302 Cergy-Pontoise}
\thanks[cnrs]{LPTM is CNRS UMR 8089}

\begin{abstract}
In random cellular systems, both observation and  maximum entropy inference give a specific form to the topological pair correlation: it is bi-affine in the cells number of edges with coefficients depending on the distance between the two cells of the pair.
Assuming this form for the pair correlations, we make explicit the conditions of statistical independence at large distance.
When, on average, the defects do not contribute, the layer population and the enclosed topological charge both increase polynomially with distance. In dimension 2, the exponent of the leading terms depend on sum rules satisfied, or not, by the maximum entropy coefficients. 
\end{abstract}

\end{frontmatter}


\section{Introduction}
Foams are random but the precise probability distributions describing their structure are still largely unknown.
Leaving aside the geometrical and energetic details, obviously dependent on the system considered, we focus on the topological characteristics, believed to be more universal.
Only a few investigations have been devoted to correlations beyond nearest neighbours.
Pair correlations at arbitrary distance were analysed by Fortes-Pina \cite{FortesPina}, Rivier {\it et al} \cite{DubertRivPe,OhlenbADRiv}, Szeto {\it et al} \cite{SzetoAsTa}, and measured in \cite{SzetoAsTa}.

Maximum entropy is one of the few methods able to predict some aspects of the probability distributions. Maximum entropy (maxent) arguments yield a specific form for the pair correlations at arbitrary distance \cite{DubertRivPe,PeshkinStRi}.

Here, we examine the asymptotic behaviour of pair correlations and how the independence of occurrences at large distances constrains the parameters in the maxent formulae.

The present analysis is mostly devoted to bi-dimensional foams, subject to extensive theoretical and experimental investigations \cite{FortesPina,DubertRivPe,OhlenbADRiv,SzetoAsTa,PeshkinStRi,Fradkov,Herdtle,glazier,Avron,stavans,Godreche,Delannay,Flyvbjerg,Mombach,elias,Earnshaw,Schliecker,WeaireHutz,Tewari,Feng,Graner},
 and where the progresses are ahead of those on 3D foams.

\section{Foam statistics}  \label{§stats}

A foam $\mathcal{F}$  divides space into $N=|\mathcal{F}|$ polygonal cells.
Here  $\mathcal{F}$ is viewed as a set of cells and $|\mathcal{F}|$
represents the number of elements in the set.

\subsection{One cell statistics}     \label{§1cell}
The local observable is $n$, the number of sides of each cell (``polygonality''), 
or the \emph{topological charge} $q=6-n$.
The fraction of $n$-sided cells is $p(n) = N(n)/N$.

The generic and physically stable vertex coordination (or degree) is $z=3$.
Euler's theorem implies:
\begin{equation}\label{eqEuler}
\langle q\rangle = \langle 6-n\rangle = \sum_n (6-n)\;p(n) = 6 \chi/N
\end{equation}
where $\chi$ characterises the embedding space: $\chi=2$ for the sphere, 0 for the torus%
\footnote{As stated here, equation (\ref{eqEuler}) only holds if the foam fills the 2D space, without boundary. Otherwise, the edges on the boundary must be counted only once in $\langle n\rangle N=2E-E_{\mathrm{boundary}}$, implying a additional boundary term in (\ref{eqEuler}) \cite{Graner}. $E$ is the total number of edges, internal and on the boundary. In particular, this is so for a bounded cluster in the plane, where $\chi=1$. In any reasonable case, $\langle q\rangle = \langle 6-n\rangle \rightarrow 0$ as $N\rightarrow\infty$. }.

In the large $N$ limit, $\langle q\rangle \rightarrow 0$.
The second moment is $\mu_2=\langle q^2\rangle = \langle (n-6)^2\rangle$.

\subsection{Two cells, correlations}     \label{§2cell}

\begin{figure}
\center \includegraphics[width=5cm]{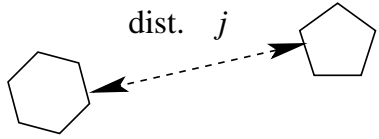}
\caption{\label{cellPair}
A pair of cells, $(k,n)=(6,5)$, at distance $j$.}
\end{figure}

\begin{figure*}
\centering  
\includegraphics[width=12cm]{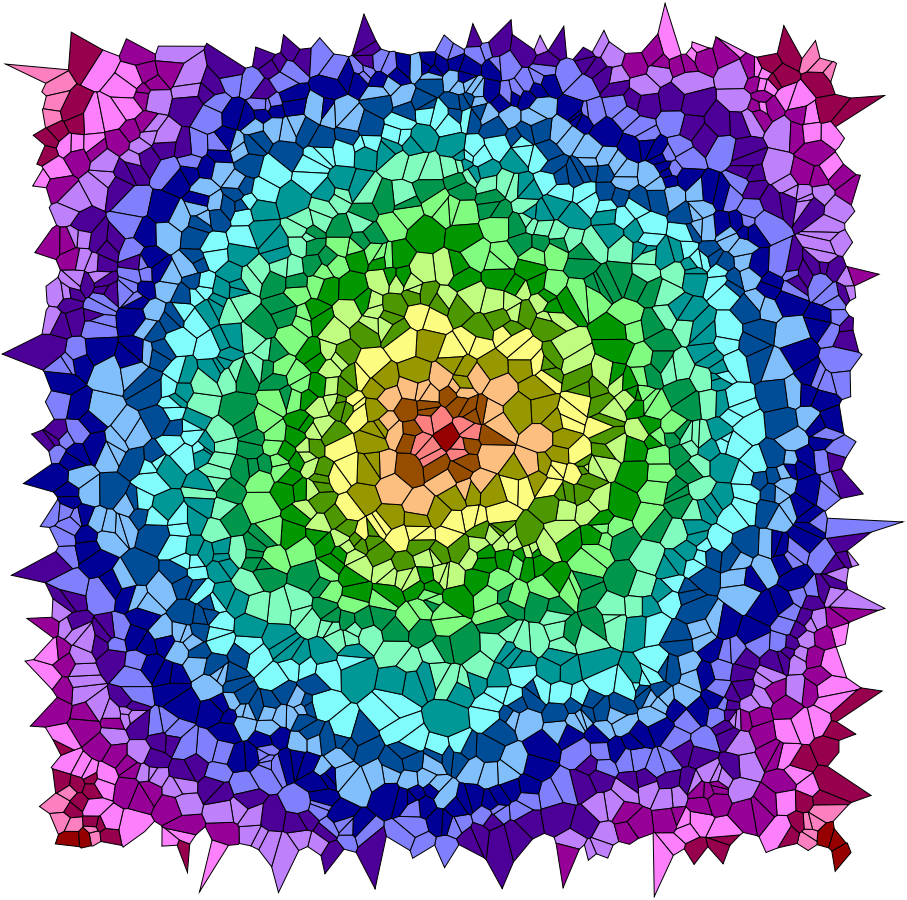}
\caption{\label{stratif}
Stratification of a Poisson-Voronoi froth into layers.}
\end{figure*}

The topological distance $j$ between cells is measured in nearest neighbour steps.
The $j$th \emph{layer} around a given cell $o$, lay$(j|o)$, is
 the set of cells at distance $j$ from $o$ (Fig. \ref{stratif}). 
It has population $K_j(o) = |$lay$(j|o)|$.
The average over $n$-sided central cells is $\langle K_j(n)\rangle$ and the overall average is $\langle\langle K_j\rangle\rangle$.

The joint distribution $p_j^{(2)}(k,n)$ --- the probability that a $(k,n)$-sided pair of cells occur at mutual distance $j$ ---
and the corresponding marginal distribution $s_j(n)=\sum_k p_j^{(2)}(k,n)$ --- the probability that a cell is at distance $j$ from an $n$-sided one\footnote{This is also the probability that a $n$-sided cell is at distance $j$ from an other one.} --- satisfy
$s_j(n) = \frac{\langle K_j(n)\rangle}{\langle\langle K_j\rangle\rangle}\; p(n)$.

The correlator $A_j(k,n)$ and correlation function $g_j(k,n)$ are defined by
\begin{eqnarray}
p_j^{(2)}(k,n)
&=& A_j(k,n) \;\frac{p(k)\:p(n)}{\langle\langle K_j\rangle\rangle} \label{eqDefA}\\
&=& g_j(k,n) \;s_j(k)\:s_j(n).     \label{eqDefg}
\end{eqnarray}
Both account for the statistical dependence of the concurrent occurrence of a $k$ and a $n$-sided cell at distance $j$. They only differ in the way they are normalised: the correlation function is 1 whereas the correlator is the mean population $\langle\langle K_j\rangle\rangle$ in independent situations.
The correlator is related to the layer population by \cite{FortesPina,DubertRivPe}
\begin{equation}\label{eqpopulation1}
\sum_k p(k) A_j(k,n) = \langle K_j(n)\rangle.
\end{equation}
A similar identity follows from counting the edges (`polygonality' $P_j$ of layer $j$) \cite{FortesPina}:
\begin{equation}\label{eqpolygony}
\sum_k k\: p(k)\: A_j(k,n) = \langle P_j(n)\rangle.
\end{equation}
The polygonality of a set is the sum of the individual polygonalities $n$.
Because charges are additive, and simply related to $n$ by $q=6-n$, it is more natural to consider the total charge\footnote{The charge of a set of cells is the sum of the individual charges $q=6-n$.}
of layer $j$, $Q(\lay(j|o))= 6 K_j(o) - P_j(o)$. Here we are getting closer to \cite{{DubertRivPe}}.
Then averaging over $n$-sided central cells $o$ yields, according to (\ref{eqpopulation1}), (\ref{eqpolygony}),
\begin{eqnarray}
\langle Q(\lay(j|n))\rangle
&=&\sum_k (6-k)\: p(k)\: A_j(k,n). \label{eqlaycharge}
\end{eqnarray}

\section{Maximum entropy}  \label{§1maxent}

Maximum entropy arguments (maxent) and the recursion relation to be described in sec. \ref{§recureq} give the following expressions, also observed in numerical simulations, \cite{DubertRivPe} 
\begin{eqnarray}\label{eqmaxent1}
A_j(k,n) &=& \sigma_j(k-6)(n-6) + a_j(n+k-12) + b_j,\\
\langle K_j(n)\rangle &=& a_j (n-6) + b_j, \label{eqKaffine}  
\end{eqnarray}
$\sigma_j, a_j, b_j$ are real parameters for $j=1,2,\ldots$. The contributions from defects have been neglected. 
In the infinite foam limit, the average of (\ref{eqKaffine}) gives $\langle\langle K_j\rangle\rangle = b_j$.

\section{Asymptotic freedom} \label{§freedom}

In normal systems of statistical physics, spatially distant events become uncorrelated.
In foams, this was first measured by \cite{SzetoAsTa}.
As $j\rightarrow \infty$,
\begin{eqnarray}    \label{eqDecorrel}
p_j^{(2)}(k,n) \rightarrow s_j(k)\,s_j(n)
& \Leftrightarrow & g_j(k,n)\rightarrow 1. 
\end{eqnarray}
With
\begin{eqnarray}\label{eqg}
g_j(k,n) &=& \frac{A_j(k,n) \langle\langle K_j\rangle\rangle }{\langle K_j(k)\rangle \langle K_j(n)\rangle}\\
&=& 1+ \left({\textstyle \frac{\sigma_j}{b_j} - (\frac{a_j}{b_j})^2} \right)(k-6)(n-6)
\end{eqnarray}
and maxent, asymptotic de-correlation holds if and only if
\begin{equation}\label{eqAsymptE}
\frac{\sigma_j}{b_j} - \left(\frac{a_j}{b_j}\right)^2 \rightarrow 0.
\end{equation}
A sufficient condition is that the ratios $\frac{\sigma_j}{b_j}\rightarrow 0$
and $\frac{a_j}{b_j}\rightarrow 0$, meaning that, in both $A_j$ and $K_j$,
the $n$ dependent terms would be dominated by the constant one ($b_j$) at large
$j$, a sensible result.
In this limit, (\ref{eqKaffine}) implies $\langle K_j(n)\rangle \rightarrow \langle\langle K_j\rangle\rangle$ and then $s_j(n) \rightarrow p(n)$; the effect of the central condition vanishes, another manifestation of asymptotic de-correlation.
In (\ref{eqDecorrel}), $s_j$ can then be replaced by $p$.

\section{Recursion equation} \label{§recureq}

The following equation was derived in \cite{FortesPina} and in \cite{DubertRivPe,RivierAs}:
\begin{equation}\label{eqRecurrenceq}
\Delta \langle K_j(n)\rangle + \pseudol q_j(n)\pseudor \langle K_j(n)\rangle
= \langle I_j(n)\rangle \simeq 0.
\end{equation}
The curly bracket quantities $\pseudol q_j(n)\pseudor$ (resp. $\pseudol m_j(n)\pseudor$) are the cellular charge (resp. sidedness) averaged over the neighbours at topological distance $j$ to an $n$-side cell. They are defined by $\pseudol q_j(n)\pseudor$ = $6-\pseudol m_j(n)\pseudor = \langle Q(\lay(j|n))\rangle/\langle K_j(n)\rangle$.
$\Delta K_j = K_{j+1} - 2 K_j + K_{j-1}$ is the discrete laplacian.

\begin{figure}[ht]
\includegraphics[width=.52\columnwidth]{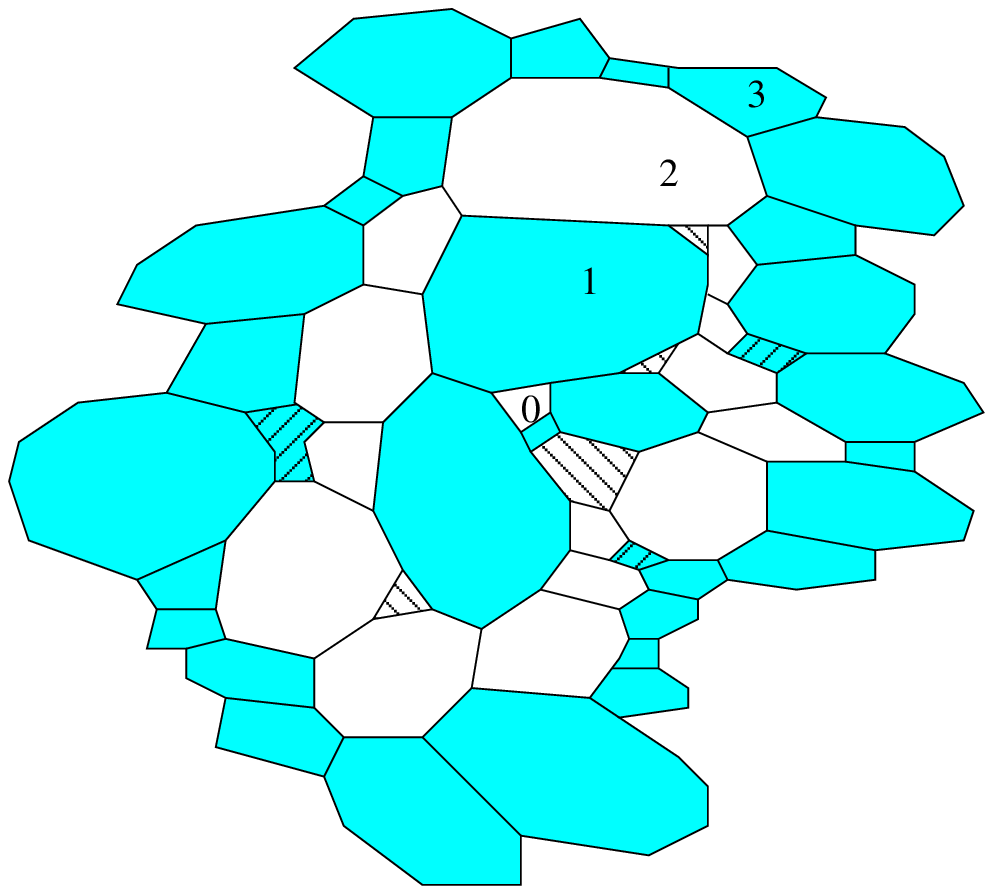}\includegraphics[width=.48\columnwidth]{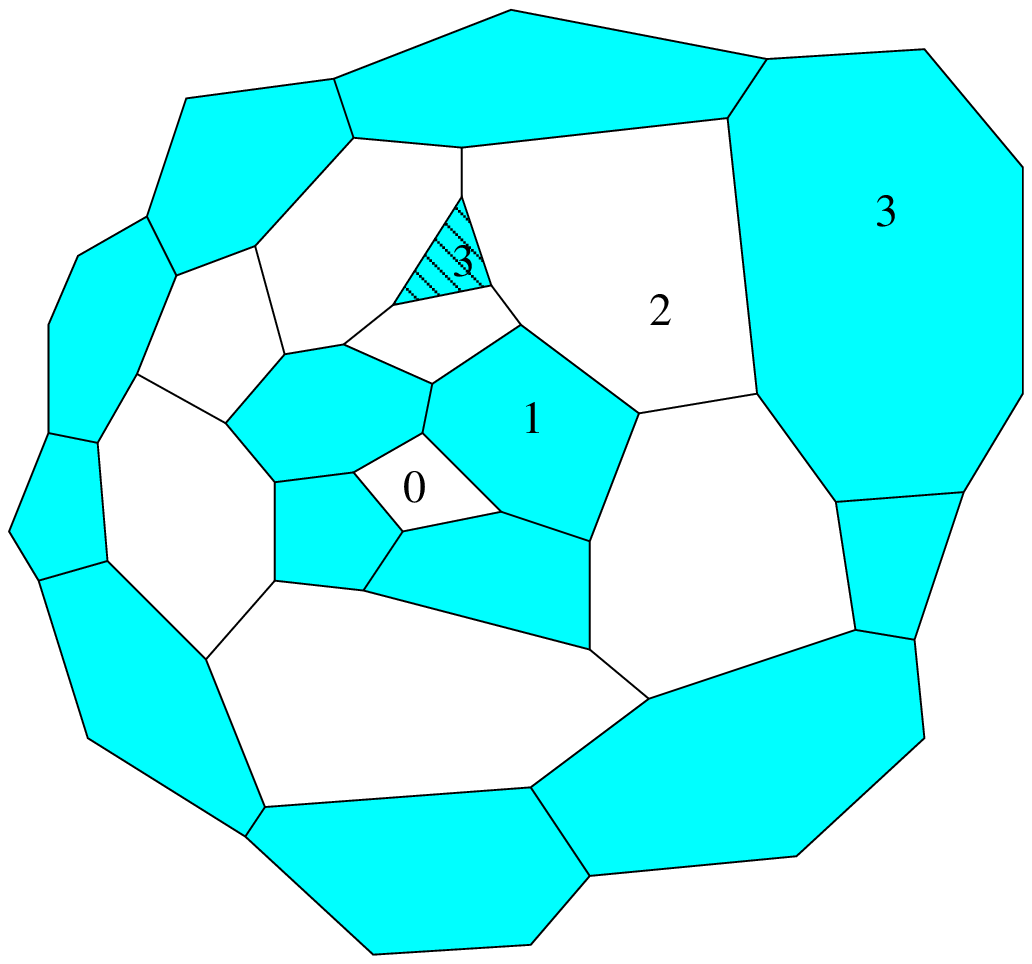}
\caption{\label{inclusions}
Defects (left) and an inclusion (right), marked hatched. 
}
\end{figure}

The right hand side
is due to the presence of \emph{defects}, the cells of layer $j$ which have no edge in common with the next layer, $j+1$ (Fig. \ref{inclusions}).
It is assumed that, {\em on average}, this contribution vanishes:
$\langle I_j(n)\rangle = 0$.

Using (\ref{eqlaycharge}) for the total charge $\langle Q(\lay(j|n))\rangle = \pseudol q_j(n)\pseudor \langle K_j(n)\rangle$ allows to put the recursion relation (\ref{eqRecurrenceq}) in the following form:
\begin{equation}\label{eqRecurrenceA}
\Delta \langle K_j(n)\rangle =
\sum_k (k-6)\: p(k)\: A_j(k,n).
\end{equation}

\section{Asymptotic behaviour and sum rules} \label{§Asympt}

With the maxent form (\ref{eqmaxent1}),(\ref{eqKaffine}) of the correlator $A_j(k,n)$ and population $\langle K_j(n)\rangle$, the recursion 
relation implies the following system, with $\alpha_j=a_j$ and $\beta_j=b_j-6a_j$:
\begin{eqnarray}\label{eqRecurabs}
\Delta \left(\begin{array}{c}\alpha_j\\\beta_j\end{array}\right)
&=& \mu_2 \left(\begin{array}{c}\sigma_j\\\alpha_j-6\, \sigma_j\end{array}\right),\\
\left(\begin{array}{c}\alpha_0\\\beta_0\end{array}\right) = 0&,&
\left(\begin{array}{c}\alpha_1\\\beta_1\end{array}\right) =
\left(\begin{array}{c}1\\0\end{array}\right).\label{eqInitcond}
\end{eqnarray}
The initial conditions follow from $\langle K_0(n)\rangle =0$ and $\langle K_1(n)\rangle = n$; $\mu_2$ was defined in sec. \ref{§1cell}.

The solutions involve $f_k = -\mu_2 \sigma_k$, conjectured to decrease fast at large $k$. 
Then, $\sum_{k=1}^{j-1} f_k \rightarrow S$ as
$j\rightarrow \infty$, so that $f/S$ defines a normalised distribution
(like probabilities except that some $f_k$ may be negative). 
Let $M_n$ be the moments of this distribution:
\begin{equation}
S= \sum_{k\geq 1} f_k, \qquad M_n = \frac{1}{S}\sum_{k\geq 1} k^n f_k.
\end{equation}

The solutions to (\ref{eqRecurabs}), (\ref{eqInitcond}) are
\begin{eqnarray}  \label{eqalphaf} 
\alpha_j &=& {\textstyle \sum_{k=1}^{j-1}} k f_k +j(1- {\textstyle \sum_{k=1}^{j-1}} f_k),\\
\beta_j  \nonumber
&=& \frac{\mu_2}{6}\left[(j+1)j(j-1)(1-{\textstyle \sum} f_k)
      + (3j^2-1){\textstyle \sum} k f_k \right. \\ & &
    - \left. 3j {\textstyle \sum} k^2 f_k + {\textstyle \sum} k^3 f_k \right]
      + 6{\textstyle \sum} (j-k) f_k,  \label{eqbetaf}
\end{eqnarray}
where the sums run from $k=1$ to $j-1$.
As $j\to\infty$, they behave like
\begin{eqnarray} \label{eqalphalim}
\alpha_j &\rightarrow& j(1-S) + S\,M_1,\\
\beta_j &\rightarrow&  \nonumber
   \frac{\mu_2}{6} S \left[(j+1)j(j-1)(1/S-1)
     + (3j^2-1) M_1 \right. \\ && 
  - \left. 3j M_2 + M_3 \right] + 6S(j - M_1), \label{eqbetalim}
\end{eqnarray}
The leading order will be lower if the $\{f_k = -\mu_2\sigma_k\}$ satisfy 
certain sum rules.
Indeed, if $S=1$, the estimates become
\begin{eqnarray} \label{eqlimb0}
\alpha_j &\rightarrow& M_1\\
\beta_j &\rightarrow \frac{\mu_2}{6}& \left[(3j^2-1) M_1 - 3j M_2 + M_3 \right] \nonumber\\
&& +\  6 (j - M_1), \label{eqlimb1}
\end{eqnarray}
so that the average population $\langle K_j(n)\rangle$ would not grow faster than $j^2$.

The so-called Euclidean scaling \cite{OhlenbADRiv,OgRiv}, $\langle K_j(n)\rangle \sim j$ in 2D, requires a second condition: 
$$M_1= -\mu_2 \sum_{k\geq 1} k\, \sigma_k = 0.$$

\section{Topological charge} \label{§Charge}
The purpose of this section is to relate the moments $S$ and $M_k$ to the average charge contained in a ball or in a layer. It does not enforce the precise values found in the previous section, as we first hoped; but it shows that the sum rules and asymptotes, for charges and populations, are compatible.

\subsection{Charge enclosed in a ball}
Let $B_j(o)$ be the union of the cells at distance at most $j$ from an arbitrarily chosen cell $o$. The total charge enclosed in the cluster $B_j(o)$ satisfies, averaging over $n$-sided origins $o$,
\begin{eqnarray} 
\langle Q(B_j)\rangle &=& \left(6 - n - \mu_2 {\textstyle \frac{j(j+1)}{2}} \right)\left[1 - \sum_{k=1}^{j} f_k\right] \nonumber\\
&+&\  \frac{\mu_2}{2} \left[ \sum_{k=1}^{j} k^2 f_k - (2j+1) \sum_{k=1}^{j} k f_k \right].
\label{eqAvChargeB}
\end{eqnarray}
where $\langle Q(B_j)\rangle$ stands for $\langle Q(B_j(n))\rangle$.
This expression follows from inserting the solutions $\alpha_j, \beta_j$ into (\ref{eqmaxent1}, \ref{eqKaffine}) and either summing the layer charges (\ref{eqlaycharge}) from 0 to $j$ or integrating the difference equation (\ref{eqRecurrenceq}).
It is consistent with (\ref{eqbetalim}).
Indeed, if the individual charges are normal random variables, averaging to zero by Euler 
equation (\ref{eqEuler}), their sum should behave like fluctuations:
\begin{equation} \label{Qpop}
|\langle Q(B_j(n))\rangle| \leq \mathrm{const}\ \langle |B_j(n)|\rangle^{1/2}.
\end{equation}
Now, if $S\neq 1$, the population $\langle |B_j(n)|\rangle \propto \sum_{l=0}^{j}\langle K_l(n)\rangle \sim j^4$ and $\langle Q(B_j(n))\rangle \sim j^2$, as in (\ref{eqAvChargeB}) without further condition.

If, on the other hand, the charge fluctuations are less, $\langle Q(B_j(n))\rangle = o(j^2)$, then $S= 1$ must hold, predicting a layer population  $\sim j^2$, according to (\ref{eqlimb1}).

A possible interpretation of $S=1$ can be deduced from (\ref{eqAvChargeB}).
Define the average $\langle\langle Q_j\rangle\rangle=\sum_n p(n) \langle Q(B_j(n))\rangle$. It coincides with the mean charge of a ball centred at a neutral cell: $\langle Q(B_j(6))\rangle$. Then, using (\ref{eqAvChargeB}), we can calculate the mean excess charge in the ball due to conditioning on $n$-sided central cells:
\begin{equation} \label{Qexcess}
\langle Q(B_j(n))\rangle - \langle\langle Q_j\rangle\rangle = (6-n)\left[1 - \sum_{k=1}^{j} f_k\right].
\end{equation}
As $j\rightarrow\infty$, this excess charge becomes $(6-n)(1-S)$; then $S=1$ means global neutrality: the central charge $q=6-n$ is exactly screened by the excess charge $(6-n)S$ in the layers around.

\subsection{Layer charge and Aboav-Weaire's law}
Inserting the maxent formula (\ref{eqKaffine}) for $\langle K_j(n)\rangle$ into (\ref{eqRecurrenceq}) and using (\ref{eqRecurabs}) gives an expression for the average charge in layer $j$:
\begin{equation}\label{eqLayQ}
\langle Q(\lay(j|n))\rangle= (n-6)f_j - \mu_2 \alpha_j,
\end{equation}
or, dividing by the mean layer population,
\begin{equation}\label{eqLayq}
\pseudol q_j(n)\pseudor = \frac{\langle Q(\lay(j|n))\rangle}{\langle K_j(n)\rangle}
= \frac{(n-6)f_j - \mu_2 \alpha_j}{\alpha_j n+\beta_j}.
\end{equation}
This is a fractional linear function of $n$ as in Aboav-Weaire's law \cite{Aboav}, which is in fact the first case, $j=1$, of this sequence labelled by $j$. Because double contacts are negligible in foams, the first layer population is just the number if sides of the central cell: $K_1(n) = \langle K_1(n)\rangle = n$ and $\pseudol q_1(n)\pseudor =  \langle q_1(n)\rangle = 6-\langle m_1(n)\rangle$, where $\langle m_1(n)\rangle$ is the mean number of sides (the polygonality) of the first neighbours of $n$-sided cells. So, with (\ref{eqInitcond}), equation (\ref{eqLayq}) for $j=1$ specialises to
\begin{equation}\label{eqLayq1}
\langle q_1(n)\rangle = \frac{(n-6)f_1 - \mu_2}{n}= f_1 - \frac{6f_1 + \mu_2}{n}.
\end{equation}
Compared to the usual form of Aboav-Weaire's law \cite{Aboav,Chiu}: 
$\langle m_1(n)\rangle = 6-a + (6a+\mu_2)/n$,
eq. (\ref{eqLayq1}) gives an interpretation of the parameter $a$ as the $n\,k$ coefficient in the pair correlator: $a = f_1 = -\mu_2 \sigma_1$.

\section{Conclusion}
To summarise, the major part of this article is a brief review of what is known, so far, on correlations in foams beyond first neighbours. This includes the definitions and sum rules in sec 2, the (bi)affine form of the correlations in sec. 3, the recursion equation in sec. 5 and the solutions in sec. 6. Except for the affine ansatz, controversially \cite{Chiu} justified by maximum entropy arguments \cite{DubertRivPe,PeshkinStRi}, and the related hypothesis that defects do not contribute on average, which we assumed from the start, our purpose has been to restrain approximations or ad hoc substitutions to a minimum.
Some approximations (factorisations, etc.) give interesting perspectives and will be treated in an other article \cite{MirOgprepa}.

The new results are mainly contained in sections \ref{§freedom}, \ref{§Asympt}, \ref{§Charge}.
First, we have shown that the maxent affine ansatz is compatible with asymptotic freedom, and how the correlation decay affects the affine coefficients.

Next, the decay of the correlations, or the rate of increase of the layer populations, at large distance are tightly related to sum rules for the coefficients $\sigma_j$ in equ. (\ref{eqmaxent1}), specifying the screening of (topological) charges. Charge neutrality means that a given charge is surrounded by a cloud of total opposite charge. In electrostatics, charge neutrality results from energy bounds and shielding the long range Coulomb field. This strong Debye screening implies subnormal charge fluctuations \cite{MartinYalcin}.

In foams, as far as we can see, no electric field constrains the topological charge.
Global neutrality is a consequence of Euler equation (\ref{eqEuler}). Neutrality seems also true at an intermediate scale. The charge fluctuations result from a strange compromise between statistical disorder and geometrical constraints \cite{magnasco}. In sec. \ref{§Asympt}, the order of the asymptotic polynomial behaviour of the layer population is related to specific sum rules satisfied by the correlation coefficients. In turn, these sum rules command the overall charge (fluctuations) in large domains.

The conclusion is that both populations and charges are consistently related. 
The next question is what do we need to get more precise, or more predictive, estimates.

\section*{Acknowledgement}
We would like to deeply thank the referees for a number of
interesting, helpful questions and comments.




\end{document}